\documentstyle[12pt]{article}
\begin{document}
\newcommand{\be}{\begin{equation}}
\newcommand{\ee}{\end{equation}}
\newcommand{\bea}{\begin{eqnarray}}  
\newcommand{\eea}{\end{eqnarray}}
\newcommand{\beas}{\begin{eqnarray*}}
\newcommand{\eeas}{\end{eqnarray*}}

\baselineskip 14 pt
\parskip 12 pt

\begin{titlepage}
\begin{flushright}
{\small hep-th/0406203}\\
\end{flushright}

\begin{center}

\vspace{2mm}

{\Large \bf Black hole thermalization rate from brane anti-brane model}

\vspace{3mm}

 Gilad Lifschytz

\vspace{1mm}

{\small \sl Department of Mathematics and Physics and CCMSC} \\
{\small \sl University of Haifa at Oranim, Tivon 36006, Israel} \\
{\small \tt giladl@research.haifa.ac.il}

\end{center}

\vskip 0.3 cm

\noindent
We develop the quasi-particle picture for Schwarzhchild and
far from extremal black holes. We show that the thermalization
equations of the black hole is recovered from the model of branes
and anti-branes. This can also be viewed as a field theory explanation of the
relationship between area and entropy for these black holes. As a by
product the annihilation rate of branes and anti-branes is computed.

\end{titlepage}

\section{Introduction}
The properties of certain near extremal black holes are encoded in the
thermodynamics of large $N$ gauge theories \cite{malda,imsy}.
 Recently it was shown that the properties of the large $N$
gauge theories at finite temperature in the large `t Hooft effective
coupling 
regime can be captured by a simple quasi-particle
description \cite{ikll1,ikll2}. In this description each
quasi-particle 
has an energy of order the temperature and thus
the number of quasi-particles is the entropy of the gauge
theory. Further
the  lifetime of the quasi-particles is of order the inverse
temperature. 
For the case
where only one type of brane is considered, the horizon size follows
from the double peaked nature of the spectral density of the 
scalars fields \cite{ikll}.

Apart from describing the equilibrium properties of the system the quasi-particle description captures the approach to equilibrium too.
It was shown that in general the membrane paradigm description of a
black hole gives the equation \cite{ikll1}
\begin{equation}
\frac{dM}{dt} \sim \frac{A}{G}T^{2}
\end{equation}
describing the response to perturbation and the approach to equilibrium of the black hole.
The quasi-particle description of the near extremal black holes gave
\begin{equation}
\frac{dE}{dt} \sim N_{qp} T^{2}
\end{equation}
where $N_{qp}$ is the number of quasi-particles.
This lead to  the identification (in these cases $dE=dM$ )of the
number of quasi particles with the area of the horizon in Planck
units, giving a simple explanation of why the entropy is proportional
to the area in Planck units. This approach was generalized to a variety
of examples including the rotating BTZ black hole.

More recently following the work in \cite{dgk}, it has been shown that
a large class of black hole (far from extremal) \cite{ghm,sp,bl,ka,lif,ks,edi}
can be described using near extremal black holes, with opposite
charges. Since the quasi-particle description gave a simple
explanation to some of the properties of  the near extremal black hole
it seems that maybe the quasi-particle picture can explain some properties of
the far from extremal black hole (including Schwarzhchild black holes).
In this note we start to develop this quasi-particle picture.

\section{Single charged black brane} 

We first want to look at the case where there are some brane and some
anti-branes of the same type. The black brane properties are described by\cite{dlp,kt}
\begin{eqnarray}
M_{bh}&=&\frac{\omega_{d+1}}{2\kappa^2}L^{p}\mu^{d}(d+1+d\sinh ^{2}\gamma)\nonumber\\
S_{bh}&=&\frac{2\pi}{\kappa^2}\omega_{d+1}L^{p}\mu^{d+1}\cosh \gamma \nonumber\\
Q_{bh}&=&\frac{1}{2\sqrt{2}\kappa}\omega_{d+1}L^{p}\mu^{d}\sinh
2\gamma \nonumber \\
T_{BH}&=&\frac{d}{4\pi \mu cosh\gamma}
\label{farext}
\end{eqnarray}

In \cite{lif} these black holes are modeled by the properties of the brane anti-brane system at finite temperature.
The field theory on the brane is described by the thermodynamical relation
\begin{eqnarray}
S&=&aE^{\lambda}\sqrt{M_{p,0}} \nonumber\\
\lambda&=&\frac{d+1}{d}-\frac{N}{2}\ \  D=p+d+3 \nonumber\\
a&=&2^{\frac{1}{d}+2}\pi \omega_{d+1}^{-\frac{1}{d}}d^{-\frac{d+1}{d}}
\lambda^{-\lambda}L^{-\frac{p}{d}}\kappa^{\frac{2}{d}} \nonumber\\
M_{p,0}&=&L^{p}\tau_{p}N 
\label{field1}
\end{eqnarray}
where $\tau_{p}$ is the tension of the brane , $N$ is the number of
branes and $E$ is the energy of the excitations.
Similarly the theory on the anti brane is described by the same relationship with $M_{p,0}$ replaced by $M_{\bar{p},0}$, and $E$ replaced by $\bar{E}$.
It was argued that for stability $E=\bar{E}$, and we will continue to assume this. Maximizing the total entropy of the brane and anti-branes subject to this constraint and at a fixed total mass and charge gave the relationship
\begin{equation}
E_{total}=2E=4\lambda\sqrt{M_{p,0}M_{\bar{p},0}}.
\label{emm}
\end{equation}

If we define $\mu$ and $\gamma$ such that they are related to the mass
and charge of the black hole as in (\ref{farext}) then we find \cite{lif}
\begin{eqnarray}
M_{p,0}=\frac{\omega_{d+1}}{2\kappa^2}L^{p}\mu^{d}\frac{d}{4}e^{2\gamma}\nonumber \\
M_{\bar{p},0}=\frac{\omega_{d+1}}{2\kappa^2}L^{p}\mu^{d}\frac{d}{4}
e^{-2\gamma}\nonumber \\
E_{tot}=\lambda\frac{\omega_{d+1}}{2\kappa^2}L^{p}\mu^{d}d
\label{sol1}
\end{eqnarray}

The theories while having the same energy have  different
temperatures, $T$ and $\bar{T}$ respectively, satisfying
\be
\frac{2}{T_{BH}}=\frac{1}{T}+\frac{1}{\bar{T}}.
\ee

Following the analysis in \cite{ikll1}, the rate at which the black
hole emits energy as seen by observers at the
stretched horizon,
 ignoring charged emission (So from now on the charge $Q$ is fixed),
is given by,
\begin{equation}
\frac{dM_{BH}}{dt}\sim \frac{A}{G}T_{BH}^{2}\sim S_{BH}T_{BH}^{2}
\label{dmbh}
\end{equation}
If we are also interested in perturbation that keep the charge
constant, then $dM=T_{BH}dS_{BH}$, and one has
\begin{equation}
\frac{dS_{BH}}{dt}\sim \frac{A}{G}T_{BH}=S_{BH}T_{BH}
\label{dsbh}
\end{equation}
Equations (\ref{dmbh}) and (\ref{dsbh}) describe how the black hole
thermalizes.

Since we claimed that the black hole described in equation
(\ref{farext}),
can be modeled by field theory living on branes and anti-branes we
should be able to reproduce equations (\ref{dmbh}) and (\ref{dsbh}),
using the out of equilibrium properties of the near extremal branes.

According to the quasi particle description of near extremal branes
 \cite{ikll,ikll1,ikll2} the thermodynamics can be approximated by free
 quasi particles. The number of quasi particles is the entropy, the
 energy of each quasi particle is around the temperature and
the life time of the quasi particles is set by the inverse temperature
of the field theory.

Now when we have a system of branes and anti-branes we have assumed
that at the stability point the theories on
them are decoupled from each other except for the constraint that the
excitation energy be the same. This is the assumption underlying the
derivation of the entropy of the far from extremal black holes. This
means that the spectral density of the theories is the same as if
the other theory does not exist. From the width of the peaks in the
spectral density, or from the scaling  of the quasi normal modes in the
background of the near extremal branes, one can read off the life time
of the quasi particle
and is thus proportional to the inverse temperature\cite{ikll2}.

Thus we can write (remembering that the number of quasi-particles is
proportional to the entropy),
\bea
\frac{dS}{dt} \sim ST \nonumber\\
\frac{d\bar{S}}{dt} \sim \bar{S}\bar{T} 
\label{st2}
\eea
where the proportionality constant is the same in both theories.
What we mean by theses equations is that the quasi particle decay and
their number changes. Off course in equilibrium quasi particles are
also created in exactly the same rate, but one can focus on their
decay and compare to the corresponding properties of the black hole.

Thus the total entropy $S_{tot}=S+\bar{S}$ satisfies
\be
\frac{dS_{tot}}{dt} \sim ST+\bar{S}\bar{T}
\label{dst}
\ee
However
\be
ST+\bar{S}\bar{T}=(S+\bar{S})\frac{2T\bar{T}}{T+\bar{T}}=S_{tot}T_{BH}
\ee
where in the last equality we have used that 
\be
\bar{S}\bar{T} =ST \sim E
\ee
We thus see that the equation (\ref{dst}) reproduces equation
(\ref{dsbh}).

While the life time of the quasi particle is the same as in the near
extremal case, the relationship
between the change in the number of quasi particles and the change in 
excitation
energy is not the same since pairs of branes and anti branes can be
created or annihilated. So while we  know the rate of quasi
particle decay this does not tell us the rate in which the total mass
of the black hole changes. Now in the gravity picture we imagine that
the black hole would slowly decay (if there was no incoming flux at
the stretched horizon to balance off the decay) and just become a
slightly less
massive black hole hence the use of (at constant charge $Q$)
\be
dM_{BH}=T_{BH}dS_{BH}.
\label{sm}
\ee

On the field theory side the brane anti brane system is only a black
hole if the energy and the number of brane and anti branes obey the
correct relationship (\ref{emm}). 
If we now allow changes in both the excitation
energy and number of branes, are we guaranteed to obey (\ref{sm}).
In particular how restrictive does the fluctuations in energy and
brane number need to be.

To try and answer this question let us look at a general change in the
energy and brane number in the field theory, using equations
(\ref{field1}) and (\ref{emm}), we can write
\be
dS =\frac{\partial S}{\partial E}dE +\frac{\partial S}{\partial
  M_{p,0}}dM_{p,0}=\frac{1}{T}dE +\frac{1}{\bar{T}}dM_{p,0}
\ee

and similarly
\be
d\bar{S}=\frac{1}{\bar{T}}d\bar{E}+\frac{1}{T}dM_{\bar {p},0}.
\ee

Now if we look for perturbation that leaves the charge constant then
$dM_{\bar {p},0}=dM_{p,0}=dM$, so we can write two equations for the
time dependence of the quantities above
\bea
\frac{dS}{dt}=\frac{1}{T}\frac{dE}{dt}
+\frac{1}{\bar{T}}\frac{dM}{dt}\nonumber \\
\frac{d\bar{S}}{dt}=\frac{1}{\bar{T}}\frac{d\bar{E}}{dt}
+\frac{1}{T}\frac{dM}{dt}
\label{oneq}
\eea
If we want to require that the change in the field theory correspond
to infinitesimal changes in a black hole configuration then one should
have (using equation (\ref{oneq}))
\be
\frac{1}{T}\frac{dE}{dt}+\frac{1}{\bar{T}}\frac{d\bar{E}}{dt}=\frac{1}{T_{BH}}
\frac{dE_{tot}}{dt}.
\label{econ}
\ee
Which is less restrictive then one would have thought, since it does
not couple $dM$ to $dE$.

Now while we have $E=\bar{E}$ for stability one does not have to
assume that $\frac{dE}{dt}=\frac{d\bar{E}}{dt}$, since for each there
is also a corresponding inflow of energy at equilibrium\footnote{Even
  though this might be true in a particular example as above}. This inflow
comes form the annihilation of the branes and anti-branes which is
converted into excitation energy just as energy excitation creates
brane anti brane pairs. Since at equilibrium the number of branes and
anti branes stays the same and the energy $E$ is constant this means
that
\be
\frac{dE}{dt}+\frac{d\bar{E}}{dt} =2\frac{dM}{dt}.
\label{equa}
\ee

We seem to have more equations then unknown.

Solving equations (\ref{oneq}), (\ref{econ}) and (\ref{equa}) and we find
\be
\frac{dE}{dt}=\frac{d\bar{E}}{dt}=\frac{dM}{dt} \sim  E T_{BH}
\label{em}
\ee
From which we can find
\be
2\frac{dE +dM}{dt}=\frac{dM_{BH}}{dt}=T_{BH}\frac{dS_{tot}}{dt} \sim S_{tot}T_{BH}^{2},
\ee
 in agreement with equation (\ref{dmbh}).

Having a solution to an over constrained system suggest that the extra
condition (\ref{equa}) which reflects the requirement of stability of
the system is somehow already incorporated in the description. This
suggest that the requirement that the energy on brane and anti branes 
be the same from which the black hole configuration arose is indeed
related to stability \cite{lif}.

It is worth presenting the decay rate for the branes in another form
\be
\frac{dM}{dt} \sim \frac{1}{a}\frac{(M\bar{M})^{\frac{3d-2}{4d}}}{\sqrt{M}+\sqrt{\bar{M}}}
\ee
which gives the decay rate as a function of the number of branes and
anti branes alone.


\section{Multi charged BH}

We now look at a more general case involving many types of branes and 
anti-branes.
The mass and entropy are given by \cite{ct,kt1},
\begin{eqnarray}
M_{bh}&=&\frac{b}{2}\mu^{D-3}(\sum_{i=1}^{n}\cosh \gamma_{i} +2\lambda)\nonumber \\
b&=&\frac{\omega_{D-2}}{2\kappa^2}(D-3)V_{p}\ \ \ ,\ 
\lambda=\frac{D-2}{D-3}-\frac{n}{2}\nonumber \\
S_{bh}&=&c\mu^{D-2}\Pi_{i=1}^{n}\cosh \gamma_{i}\nonumber \\
c&=&\frac{4\pi b}{D-3}.\nonumber\\
T_{BH}&=&\frac{d}{4\pi \mu \Pi_{i=1}^{n} \cosh\gamma_{i}}
\label{farext2}
\end{eqnarray}

The field theory on the branes anti-branes configuration can be
summarized by \cite{lif},
\begin{eqnarray}
M_{f}=\sum_{i}^{n}(M_{p_{i},0}+M_{\bar{p}_{i},0})+E_{tot} \nonumber \\
S_{f}=\tilde{a}(\frac{E_{tot}}{2^{n}})^{\lambda}\Pi_{i=1}^{N}(\sqrt{M_{p_{i},0}}+\sqrt{M_{\bar{p}_{i},0})}.\nonumber \\
\tilde{a}=c(b\lambda)^{-\lambda}b^{-n/2}
\label{field2}
\end{eqnarray}
Where each of the $2^{n}$  configuration of branes and anti-branes has
the same energy of excitations, but different temperatures.
The different temperatures are related to the black hole temperature
by
\be
\frac{2^{n}}{T_{BH}}=\sum_{i=1}^{2^{n}}\frac{1}{T_{i}}
\label{temp}
\ee

Maximizing the entropy for fixed charge and mass gives the
relationship
\begin{equation}
E_{tot}=4\lambda \sqrt{M_{p_{i},0}M_{\bar{p}_{i},0}},
\end{equation}
for all $i$.

In terms of $\mu$ and $\gamma$ one has,
\begin{eqnarray}
M_{p_{i},0}=\frac{b}{4}\mu^{D-3} e^{2\gamma_{i}}\nonumber \\
M_{\bar{p}_{i},0}=\frac{b}{4}\mu^{D-3} e^{-2\gamma_{i}}\nonumber \\
E_{tot}= b\lambda \mu^{D-3} 
\label{sol3}
\end{eqnarray}

We now want to recover (\ref{dsbh}) and (\ref{dmbh}) from a
quasi-particle description.

As before we assume each of the $2^{n}$ field theories is described by
quasi-particles with energy of order the temperature and life time of
order the inverse temperature.

\be
\frac{dS_{i}}{dt} \sim S_{i}T_{i} \ \ \ \ i=1 \cdots 2^{n}
\ee

The time derivative of the total entropy is then
\be
\frac{dS_{tot}}{dt}\sim \sum_{i} S_{i}T_{i}\sim S_{tot}T_{BH}
\label{dtsgen}
\ee
where in the last equality we have used equation (\ref{temp}) and that
all the $S_{i}T_{i}$ are equal for all $i$, due to the equality of the
energy of excitation on all $2^{n}$ brane configurations.

We now want to compute the individual rates of the branes decay and
reproduce equation (\ref{dmbh}).
When the number of quasi particles changes this can come about by
either a change in the energy or a change in the number of branes. But
each of the $2^n$ field theories live on some of the same branes thus
they affect each other. While they can not affect the life time of the
quasi-particles they can affect the distribution of the available
energy
between branes and excitations. To be concrete let us 
look at the example of $n=2$

\subsection{The $n=2$ case}
In this case we have for the entropy
\be
S_{tot}=aE^{\lambda}(\sqrt{M_{1}}\sqrt{\bar{M}_{2}}+
\sqrt{M_{1}}\sqrt{M_{2}}+\sqrt{\bar{M}_{1}}\sqrt{M_{2}}+\sqrt{\bar{M}_{1}}\sqrt{\bar{M}_{2}})
\ee
and we can write it as 
\be
S_{tot}=S_{1}+S_{2}+S_{3}+S_{4},
\ee
respectively.
The equations for the rate of change are
\bea
\dot{S}_{1}=\frac{1}{T_{1}}\dot{E_{1}}+\frac{1}{2T_{4}}\dot{M}_{1}+\frac{1}{2T_{2}}\dot{\bar{M}}_{2}\nonumber\\
\dot{S}_{2}=\frac{1}{T_{2}}\dot{E_{2}}+\frac{1}{2T_{3}}\dot{M}_{1}+\frac{1}{2T_{1}}\dot{M}_{2}\nonumber\\
\dot{S}_{3}=\frac{1}{T_{3}}\dot{E_{3}}+\frac{1}{2T_{2}}\dot{\bar{M}}_{1}+\frac{1}{2T_{4}}\dot{M}_{2}\nonumber\\
\dot{S}_{4}=\frac{1}{T_{4}}\dot{E_{4}}+\frac{1}{2T_{1}}\dot{\bar{M}}_{1}+\frac{1}{2T_{3}}\dot{\bar{M}}_{2}
\label{eqn2}
\eea
Where we have taken into account the effect of the field theories on
each other by equating the decay rate of $M_{i}$ on each field theory
that live on these branes.
Now we have eight unknown but only four equations. Four more equation
are given by 
the equality
of the rate of energy decay to brane decay, charge conservation, and
the analog of equation (\ref{econ}) i.e
\be
\sum_{i}(\frac{\dot{E}_{i}}{T_{i}})=\frac{1}{T_{BH}}\dot{E}_{tot}
\label{econg}
\ee

 These together give eight equations which can be solved to give the
 individual decay rates. 

The charge conservation equation plus  (\ref{econg}) and
the structure of (\ref{eqn2}) give 
\be
dS_{tot}=\frac{1}{T_{BH}}(dE_{tot}+2dM_{1}+2dM_{2})
\ee
ensuring equation (\ref{dmbh}).


\end{document}